\documentstyle[preprint,aps]{revtex}

\begin{document}

\preprint{EHU-FT/9905}

\draft

\title{COLLISION TERMS  FROM FLUCTUATIONS IN THE \\ 
HTL THEORY FOR THE QUARK-GLUON PLASMA} 
\author{M. A. Valle Basagoiti \thanks{\tt wtpvabam@lg.ehu.es}}
\address{Departamento de F\'\i sica Te\'orica, \\ 
Universidad del  Pa\'\i s Vasco, Apartado 644, E-48080 Bilbao, Spain}
\date{March 23, 1999}

\maketitle

\setcounter{page}{0}

\thispagestyle{empty}

\begin{abstract}
Starting from the kinetic formulation of the hard thermal loop effective 
theory, we have (re)derived the collision terms for soft modes of order 
$g^2 T \log(1/g)$ by averaging the statistical fluctuations in the 
plasma. 
\end{abstract}

\vfill

\noindent

\eject

Over the past year, significant progress has been achieved in understanding 
the effective dynamics of soft excitations $p \sim g^2 T$ in the non abelian 
quark-gluon plasma at high temperature. B\"odeker \cite{bo} has derived 
a diffusive field theory that involves only gauge fields with dynamics 
governed by a Langevin equation, in which the gaussian noise is
only parametrised by the color conductivity, which shows a logarithmic
dependence on $g$. As an application, the numerical value of the leading log
behavior of the sphaleron rate, which measures 
baryon number violation in the hot standard model, has been 
computed \cite{moore}. 

The crucial point in B\"odeker's original derivation is the way in which the 
integration  
of hard $\sim T$ and semi-hard $\sim g T$ scales is done. The contribution of 
hard fields is easy to find. It is encoded in the local formulation of the 
hard-thermal-loop 
effective action \cite{blaizot,kelly} leading to a closed set of 
collisionless kinetic equations and field equations for the semi-hard modes. 
The remaining integration of the semi-hard scales turns out to be  
a difficult task. B\"odecker's main result is   
an effective kinetic equation with collision 
terms arising from the averaging over semi-hard fields. This 
equation has also been proposed by Arnold, Son and Yaffe \cite{arnold} by 
analyzing the scattering processes between hard particles in the plasma.  
Very recently, a Boltzmann equation has been rigorously derived 
by Blaizot and Iancu \cite{blaizot2},  
starting from the 
Dyson-Schwinger equations. For the QCD plasma, it coincides with the one 
previously obtained in Refs. \cite{bo,arnold} 

The main purpose of this paper is to give an alternative derivation 
{\it\`a la} Balescu-Lenard \cite{selikhov} of 
the collision terms for the 
longwavelength color deviations of equilibrium.  The starting point is the
set of  ``microscopic'' dynamical equations coming from 
the hard-thermal loop effective action describing the evolution of the 
collisionless plasma. In this way, collision terms arise as 
statistical averages of correlators of the plasma fluctuations at 
equilibrium \cite{klim,lif}. A similar derivation 
starting from classical transport theory has been given
recently by Litim and Manuel \cite{litim}. 
It should be emphasized that the main point here, is to  
provide a derivation directly from the 
HTL effective action.
As shown in Ref. \cite{litim}, a strictly classical derivation
leads to a different value for the  
color conductivity because the value of the  Debye mass computed 
from a Maxwell-Boltzmann distribution is different.    
 
Let us outline our calculations. Our construction 
relies on the local formulation of the HTL theory 
\cite{blaizot}, which involves a set of coupled equations of motion for the 
mean fields and their induced current. The mean fields $A^a_{\mu}(x)$ 
satisfy the Yang-Mills equations with an induced current on the right hand 
side: 
 \begin{equation}
 [D_{\nu}, F^{\nu \mu}]_a = j^\mu_{a}\,,
 \end{equation}
where $ D_{\nu} O(x) = \partial_{\nu} O(x) + i g [A_{\nu}, O(x)] $, 
$A_{\nu} = A^a_{\nu} t^a$ and $F_{\mu \nu} = [D_{\mu},D_{\nu}]/(i g)$. 
(The generators of the $SU(N_{c})$ gauge group in the 
fundamental representation are 
denoted by $t^a$; they satisfy $[t^a, t^b] = i f^{a b c} t^c$ and 
$\mbox{tr}(t^a t^b) = \delta^{a b}/2$.) Furthermore, the induced 
color current $j^\mu$ is related to the  color fluctuations in the 
gluon color densities $\delta N({\bf p}, x) = 
\delta N^a({\bf p}, x) t^a$ and the quark  
color densities $\delta n_{\pm}({\bf p}, x) = 
\delta n^a_{\pm}({\bf p}, x) t^a$ by
 \begin{equation}
 j^\mu (x) = g \int \frac{d^3 p}{(2 \pi)^3} v^\mu \, 
  \left[ 2 N_{c} \delta N({\bf p}, x) + N_{f} \delta n_{+}({\bf p}, x) -
     N_{f}\delta n_{-}({\bf p}, x) \right]\,.
 \end{equation}
 In this equation, $v^{\mu} \equiv (1,{\bf v})$, 
 ${\bf v} \equiv {\bf p}/p$ and $N_{f}$ is the number of quark flavors. The 
 system is closed by the non-abelian generalization of the Vlasov equations:
 
 \begin{eqnarray}
  \left[v \cdot D, \delta N({\bf p}, x)\right] &=& 
            -g \, {\bf v} \cdot {\bf E}(x) 
  	                            \frac{{\mbox d} N(p)}{{\mbox d} p}\,,
  	                            \label{kin1} \\ 
  \left[v \cdot D, \delta n_{\pm}({\bf p}, x)\right] &=& 
            \mp g \,{\bf v} \cdot {\bf E}(x) 
  	                            \frac{{\mbox d} n(p)}{{\mbox d} p}\,,
  	                            \label{kin2}  
\end{eqnarray}
where $N(p) = 1/(\exp(\beta p) -1)$, $n(p) = 1/(\exp(\beta p) +1)$ and 
$E^i_{a} \equiv F^{i 0}_{a}$. The current $j^\mu$ is covariantly conserved. 

These equations contain the semi-hard degrees of freedom. 
In order to 
obtain a kinetic description for the soft modes, we decompose the 
fields into mean values and fluctuations:
\begin{eqnarray}
A^\mu_{a} &\rightarrow& \overline{A}^\mu_{a} + \delta A^\mu_{a} \,, \\ 
\delta N  &\rightarrow& \overline{\delta N} + \delta N \,, \\
\delta n_{\pm}  &\rightarrow& \overline{\delta n}_{\pm} + \delta n_{\pm} \,.
\end{eqnarray} 
and average the previous microscopic equations. As a result, one obtains 
a set of equations for the average soft gauge fields $\overline{A}^a_{\mu}(x)$, 
and for the mean values of the distribution functions 
$\overline{\delta N}$, $\overline{\delta n}_\pm$:  
\begin{eqnarray}
   	\left[\overline{D}_{\nu}, \overline{F}^{\nu \mu}\right]_a 
   	= \overline{j}^\mu_{a} &+&   \mbox{non-linear terms},  \\
   	\left[v \cdot \overline{D}, \overline{\delta N}({\bf p}, x)\right] 
   	+ g \, {\bf v} \cdot {<\bf E}(x)> 
  	                            \frac{{\mbox d} N(p)}{{\mbox d} p} &=&
  	-i g \, v \cdot <\left[ \delta A, \delta N \right]>\,, 
  	\label{kinetic1} \\
   	\left[v \cdot \overline{D}, \overline{\delta n}_\pm({\bf p}, x)\right] 
   	\pm g \,{\bf v} \cdot {<\bf E}(x)> 
  	                            \frac{{\mbox d} n(p)}{{\mbox d} p} &=&
  	-i g \, v \cdot <\left[ \delta A, \delta n_{\pm} \right]>\,, 
  	\label{kinetic2}
\end{eqnarray}
where $\overline{D}_{\nu} = \partial_{\nu} + 
i g [\overline{A}_{\nu}, \cdot]$ and $<{\bf E}^i> = 
\overline{F}^{i 0}$. Thus, the set of equations above 
includes the ``collision 
integrals'' which are determined by the correlators of the fluctuations. A 
noteworthy feature of the collision terms is their non-Abelian origin. 
A similar analysis starting with the abelian Vlasov-Maxwell equations 
for a QED plasma leads 
to no collision terms because of the linearity underlying this case. 

The main difficulty in computing the collision terms  is caused by  
the nonlinearity of the equations for the fluctuations. However, the situation 
becomes significantly simpler for  a plasma at high temperature with 
the coupling constant assumed to be small. In this case, we can 
neglect terms of higher order in the fluctuations \cite{klim} and the 
following linearized equations for the fluctuations can be used:
 \begin{eqnarray}
   	\left[\overline{D}_{\nu}, \delta F^{\nu \mu}\right]_a + 
   	 i g \left[\delta A_{\nu}, \overline{F}^{\nu \mu}\right]_a
   	&=& \delta j^\mu_{a} ,  \\
   	\left[v \cdot \overline{D}, \delta N({\bf p}, x)\right] + 
   	 i g \, v \cdot \left[\delta A, \overline{\delta N}\right] 
   	&=& -g \, {\bf v} \cdot \delta {\bf E}(x) 
  	                            \frac{{\mbox d} N(p)}{{\mbox d} p}\,, \\
   	\left[v \cdot \overline{D}, \delta n_{\pm}({\bf p}, x)\right] + 
   	 i g \, v \cdot \left[\delta A, \overline{\delta n}_{\pm}\right] 
   	&=& \mp g \, {\bf v} \cdot \delta {\bf E}(x) 
  	                            \frac{{\mbox d} n(p)}{{\mbox d} p}\,,
\end{eqnarray} 
where $\delta F_{\mu \nu} =  \overline{D}_{\mu} \delta A_{\nu} - 
\overline{D}_{\nu} \delta A_{\mu}$ and 
${\delta E^i} = \delta F^{i 0}$. In order to obtain the explicit 
form of the kinetic equations we are looking for, 
it will suffice to linearize the collision terms in the 
mean values of the distribution functions, since the left-hand sides of 
Eqs.~(\ref{kinetic1})--(\ref{kinetic2}) are already 
linear functions of them. Thus, the task is to compute the fluctuations up to 
first order in $\overline{\delta N}$, $\overline{\delta n}_\pm$ and 
zero order in $\overline{A}$. Moreover, I will only consider mean 
color distributions with no dependence in $(t, \bf{x})$, which means that the 
results to be derived must be considered as zero-order terms in some
derivative expansion.    

In order to implement the suggested programme, it is customary to use 
Laplace transforms, such as
\begin{eqnarray}
\delta N_{a}(\omega, {\bf k}; {\bf p}) &=& 
\int d^3 x \int_{0}^\infty d t\, e^{-i ({\bf k} \cdot {\bf x} - \omega t)} 
 \delta N_{a}(t, {\bf x}; {\bf p})\,, \\
\delta A^i_{a}(\omega, {\bf k}) &=& 
\int d^3 x \int_{0}^\infty d t\, e^{-i ({\bf k} \cdot {\bf x} - \omega t)} 
 \delta A^i_{a}(t, {\bf x})\,, 
\end{eqnarray}
In terms of these, when the Coulomb gauge is used,
the first order correlator can be written as 
\begin{eqnarray}
<\delta A^i_{a}({\omega,\bf k}&)& 
 \delta N_{b}(\omega^\prime,{\bf k}^\prime; {\bf p})>^{(1)} = 
 \frac{i g f^{b m n} \overline{\delta N}^n({\bf p}) \hat{p}^j}
 {\omega - {\bf k}\cdot \hat{\bf p}}\, 
 <\delta A^i_{a}({\omega,\bf k}) 
 \delta A^j_{m}(\omega^\prime,{\bf k}^\prime)> \nonumber \\ 
 & & - \frac{i g^2 f^{a m n}}{k^2 -\omega^2 \epsilon_{\bot}(\omega,k)}
<\delta A^k_{m}({\omega,\bf k}) 
 \delta N_{b}(\omega^\prime,{\bf k}^\prime; {\bf p})>
 \left(\delta^{i j} - \hat{k}^i \hat{k}^j \right)  \nonumber \\ 
 & & \times \int \frac{d^3 p^\prime}{(2 \pi)^3} 
  \frac{\hat{p}^{\prime j} \hat{p}^{\prime k}}
  {\omega - {\bf k}\cdot \hat{\bf p}^\prime}
  \left( 2 N_{c} \overline{\delta N}^n({\bf p}^{\prime}) + 
           N_{f} \overline{\delta n}_{+}^n({\bf p}^{\prime}) - 
           N_{f} \overline{\delta n}_{-}^n({\bf p}^{\prime}) \right)\,, 
           \label{cor1}
\end{eqnarray}
with a similar expression for $<\delta A^i_{a}({\omega,\bf k}) 
 \delta n^b_{\pm}(\omega^\prime,{\bf k}^\prime; {\bf p})>^{(1)}$. In this 
equation, $\epsilon_{\bot}(\omega,k)$ represents the transverse part 
of the dielectric function, defined by 
\begin{eqnarray}
\epsilon^{i j}(\omega, k) &=& \epsilon_{\|}(\omega, k) 
\hat{k}^i \hat{k}^j + \epsilon_{\bot}(\omega, k) 
(\delta^{i j} - \hat{k}^i \hat{k}^j) \nonumber \\ 
&=& \delta^{i j} + \frac{2 g^2}{\omega} 
\int \frac{d^3 p}{(2 \pi)^3} \frac{\hat{p}^i \hat{p}^j}
{\omega - {\bf k}\cdot \hat{\bf p}} \left( N_{c} N^\prime(p) + 
 N_{f}\, n^\prime(p) \right)\,.
 \label{die}
\end{eqnarray}             
Eqs.~(\ref{cor1}) and (\ref{die}) deserve some comments: 

{\bf i)} 
The prescriptions 
$\omega \rightarrow  \omega + i 0^+$, 
$\omega^\prime \rightarrow  \omega ^\prime + i 0^+$ are implicit in all
denominators because the (double) Mellin's inversion formula for 
Laplace transforms requires that all singularities  lie in the 
lower half $\omega$, $\omega^\prime$ planes. 

{\bf ii)} the  correlators over the initial conditions 
for $t=0$, 
 \begin{eqnarray}
 	<\delta N_{a}({\bf k};{\bf p}) 
 	 \delta N_{b}({\bf k}^\prime; {\bf p}^\prime)>&=&  
 	 (2 \pi)^3 \delta({\bf k}+{\bf k}^\prime) \delta^{a b}\left(
 	 (2 \pi)^3 \delta({\bf p}-{\bf p}^\prime) f_{N}(p) + 
 	 \mu_{N}({\bf k};{\bf p}, {\bf p}^\prime) \right)\,,
 	\label{ini1}  \\
    <\delta n^a_{\pm}({\bf k}; {\bf p}) 
 	 \delta n^b_{\pm}({\bf k}^\prime; {\bf p}^\prime)>& = & 
 	 (2 \pi)^3 \delta({\bf k}+{\bf k}^\prime)  \delta^{a b}\left(
 	 (2 \pi)^3 \delta({\bf p}-{\bf p}^\prime) f_{n}(p) + 
 	 \mu_{n}({\bf k};{\bf p}, {\bf p}^\prime) \right)\,,
 	\label{ini2}  \\
    <\delta A^i_{a}({\bf k}) 
 	 \delta A^j_{b}({\bf k}^\prime)>& = & 
 	 (2 \pi)^3 \delta({\bf k}+{\bf k}^\prime)  \delta^{a b} 
 	 \left(\delta^{i j} - \hat{k}^i \hat{k}^j \right) 
 	 \mu_{A}({\bf k})\,,
 	\label{ini3}
\end{eqnarray}
 have played an important role, since in Eqs.~(\ref{cor1}) and (\ref{die}) 
 only  
 potential non-zero contributions to the large time behavior 
 have been written. This is so 
 because, as proven in \cite{lif},  when     
 $f_{N,n}(p)$, $\mu_{N,n}({\bf k};{\bf p}, {\bf p}^\prime) )$ 
 and $\mu_{A}({\bf k})$ are smooth functions of the momenta,  
 the only contributions giving a non damped function 
 of $t$ and $t^\prime$ come from the functions  $f_{N,n}(p)$. Thus, 
 the initial conditions for $\delta A$, $\mbox{d}\delta A/\mbox{d} t$ and 
 the specific form of the two-particle correlation functions  do not matter.
 All the information about the large time behavior is 
 included in $f_{N,n}(p)$ whose 
 explicit form can be determined as follows. After substitution 
 of Eqs.~(\ref{ini1})--(\ref{ini3}) into the 
 appropriate correlators we find 
 \begin{eqnarray}
 <\delta A^i_{a}&(&\omega, {\bf k}) 
  \delta A^j_{b}(\omega^\prime, {\bf k}^\prime)> \equiv 
  (2 \pi)^4 \delta({\bf k}+{\bf k}^\prime) 
           \delta(\omega+\omega^\prime) \delta^{a b} 
 	 \left(\delta^{i j} - \hat{k}^i \hat{k}^j \right) S(\omega,k) \nonumber \\ 
 &=& (2 \pi)^4 \delta({\bf k}+{\bf k}^\prime) 
           \delta(\omega+\omega^\prime) \delta^{a b} 
 	 \left(\delta^{i j} - \hat{k}^i \hat{k}^j \right) \\ 
 & & \times \frac{g^2 (1-\omega^2/k^2)}
 {|-\omega^2 \epsilon_{\bot}(\omega,k)+k^2|^2}\, 
 \frac{1}{4 \pi k}  
 \int_{0}^\infty d p\,p^2 \left(4 N_{c}^2 f_N(p) + 2 N_{f}^2 f_n(p)\right)\,. 
 \end{eqnarray} 
 Then, a comparison with the prediction of the fluctuation-dissipation theorem
 relating the correlator of field fluctuations with 
 the imaginary part of the retarded propagator, 
 \begin{equation}
	S(\omega, k) = (1 - e^{-\beta \omega})^{-1} 2\, \mbox{Im}\, \frac {1}
	{k^2 -(\omega+i 0^+)^2 \epsilon_{\bot}(\omega+i 0^+,k)} \,,
 \end{equation}
 reveals the right values for the initial correlators,
 \begin{eqnarray}
 f_{N}(p) &=& N_{c}^{-1}\,N(p) (1+N(p)) = 
  -N_{c}^{-1} T N^\prime(p)
 \label{corr1} \,, \\ 
 f_{n}(p) &=& 2 N_{f}^{-1} n(p) (1-n(p)) = 
  -2 N_{f}^{-1} T n^\prime(p) \,,
 \label{corr2} 
 \end{eqnarray}
where the factors $N_{c}^{-1}$ and $2 N_f^{-1}$  
compensate similar factors in the color current. These results for the 
initial correlators could also be established from the general properties of 
the fluctuations of an ideal gas \cite{sta}. Up to the color 
and flavor factors, Eqs.~(\ref{corr1}) and 
(\ref{corr2}) are exactly the equations for the mean square of the 
fluctuations presented in Ref. \cite{sta}\footnote{Specifically, Eqs. 
(113,3) and (113,4) of Ref. \cite{sta}.}.    

Now, following the Ref. \cite{lif}, other correlators 
can be easily computed. For example, at leading order in $g$ we have 
\begin{eqnarray}
<\delta A^i_{a}(\omega,{\bf k}) 
 \delta N_{b}(\omega^\prime, {\bf k}^\prime; {\bf p})> &=&  
 (2 \pi)^4 \delta({\bf k}+{\bf k}^\prime) 
           \delta(\omega+\omega^\prime)\,
 4 \pi g \, \delta^{a b} 
  \left(\delta^{i j} - \hat{k}^i \hat{k}^j \right) \hat{p}^j \nonumber \\  
 & & \times   \frac{N(p) (1+N(p))}{k^2 -\omega^2 \epsilon_{\bot}(\omega,k)}\,
     \delta(\omega - {\bf k} \cdot \hat{\bf p})\,, \\ 
<\delta A^i_{a}({\omega,\bf k}) 
 \delta n^{b}_{\pm}(\omega^\prime,{\bf k}^\prime; {\bf p})> &=&  
 \pm (2 \pi)^4 \delta({\bf k}+{\bf k}^\prime) 
           \delta(\omega+\omega^\prime)\, 
 2 \pi g \, \delta^{a b}
  \left(\delta^{i j} - \hat{k}^i \hat{k}^j \right) \hat{p}^j \nonumber \\  
 & & \times   \frac{n(p) (1-n(p))}{k^2 -\omega^2 \epsilon_{\bot}(\omega,k)}\,
     \delta(\omega - {\bf k} \cdot \hat{\bf p})\,. 
\end{eqnarray}

{\bf iii)} The dominant scattering processes in the  collisions 
correspond to $t$-channel exchange of 
quasistatic magnetic gauge bosons, which means that the longitudinal part of 
the interaction can be ignored.  
Also, as the relevant regime to be 
considered in transverse propagators is $\omega \ll k$, we can make the 
following replacements
\begin{eqnarray} 
\epsilon_{\bot}(\omega,k)&\simeq&\frac{3\pi\,i\,\omega_{p}^2}{4\omega k}\,,\\
 S(\omega,k) &\simeq& \frac{2 \pi}{k^2 \beta}\, \delta(\omega)\,, \\ 
\frac{1}{(-k^2 + \omega^2 \epsilon_{\bot}(\omega,k))^2} &\simeq& 
\frac{2}{3 k \omega_{p}^2} \, \delta(\omega)\,,
\end{eqnarray}
with the plasma frequency $\omega_{p}^2 = g^2 (2 N_{c} + N_{f}) T^2/18$.

Finally, the collision terms easily emerge from all of this by 
integration over $(\omega,{\bf k})$. Their explicit 
form turns out to be 
\begin{eqnarray}
-i g \, v \cdot &<&\left[ \delta A, \delta N \right]^a> = -\gamma_{g}
 \left[ \overline{\delta N}^a(x, {\bf p}) + 
 \frac{4 g^2}{3 \pi \omega_{p}^2}\, N^\prime(p) \nonumber \right. \\
   & & \times \left.  \int \frac{d^3 p^\prime}{(2 \pi)^3} 
 \frac{(\hat{{\bf p}}\cdot\hat{{\bf p}}^\prime)^2}
  {\sqrt{1 - (\hat{{\bf p}}\cdot\hat{{\bf p}}^\prime)^2}}  
  \left( 2 N_{c} \overline{\delta N}^a(x, {\bf p}^\prime) + 
      N_{f} \overline{\delta n}_{+}^a(x, {\bf p}^\prime) - 
      N_{f} \overline{\delta n}_{-}^a(x, {\bf p}^\prime \right) \right]\,, \\ 
-i g \, v \cdot &<&\left[ \delta A, \delta n_{\pm} \right]^a> = -\gamma_{g}
 \left[ \overline{\delta n}_{\pm}^a(x, {\bf p})  \pm  
 \frac{4 g^2}{3 \pi \omega_{p}^2}\, n^\prime(p) \nonumber \right. \\
   & & \times \left.  \int \frac{d^3 p^\prime}{(2 \pi)^3} 
 \frac{(\hat{{\bf p}}\cdot\hat{{\bf p}}^\prime)^2}
  {\sqrt{1 - (\hat{{\bf p}}\cdot\hat{{\bf p}}^\prime)^2}}  
  \left( 2 N_{c} \overline{\delta N}^a(x, {\bf p}^\prime) + 
      N_{f} \overline{\delta n}_{+}^a(x, {\bf p}^\prime) - 
      N_{f} \overline{\delta n}_{-}^a(x, {\bf p}^\prime \right) \right]\,,  
\end{eqnarray}
with the damping rate $\gamma_{g}$ for the hard thermal gauge bosons 
\begin{equation}
\gamma_{g} = \frac{g^2 N_c T}{4 \pi} \log(1/g)\,. 
\end{equation}
If we combine the different color distributions functions in the 
form \cite{blaizot,arnold} 
\begin{equation}
W^a(x, \hat{{\bf p}}) = \frac{1}{3 \omega_{p}^2} 
 \int \frac{dp\, 4 \pi p^2}{(2 \pi)^3}\,
\left( 2 N_{c} \overline{\delta N}^a(x, {\bf p}) + 
      N_{f} \overline{\delta n}_{+}^a(x, {\bf p}) - 
      N_{f} \overline{\delta n}_{-}^a(x, {\bf p}) \right) \,,
\end{equation}
we obtain 
\begin{eqnarray}
\left[v \cdot \overline{D}, W(x, \hat{{\bf p}}) \right]^a 
   	&-& {\bf v} \cdot {<\bf E}^a(x)> = -\delta C[W]^a\,, \\
\delta C[W]^a &=&  \gamma_{g} \left( W^a(x, \hat{{\bf p}}) - \frac{4}{\pi} 
\int d\Omega_{p^{\prime}} \frac{(\hat{{\bf p}}\cdot\hat{{\bf p}}^\prime)^2}
  {\sqrt{1 - (\hat{{\bf p}}\cdot\hat{{\bf p}}^\prime)^2}} 
  W^a(x, \hat{{\bf p}^\prime}) \right)   \, ,
\end{eqnarray}
from which 
the color conductivity $\sigma = \omega_{p}^2/\gamma_{g}$ 
follows \cite{bo,arnold,sel2}. 

In conclusion, I have derived from the microscopic HTL equations 
the explicit 
form of the collision terms recently proposed by 
Arnold, Son and Yaffe \cite{arnold} 
and Blaizot and Iancu \cite{blaizot2} which are 
relevant for B\"odeker's effective theory. 
As the authors of the Ref. \cite{litim} have pointed out in a strictly 
classical context,  
this approach is simple. It is based on the properties of the 
plasma fluctuations 
at equilibrium. In this sense, this approach would be closely connected 
with linear response theory based on ``Kubo'' formulas expressing 
the transport coefficients.

 \subsection*{Acknowledgements}
 
 I am grateful to A. Ach\'ucarro, I. L. Egusquiza, M. A. Go\~ni 
 and J. L. Ma\~nes 
 for some discussions. This work has been supported in part by 
 a University of the Basque Country Grant UPV 063.310-EB 187/98.


\begin{references}

\bibitem{bo} D. B\"odeker, Phys. Lett. {\bf B426}, 351 (1998).

\bibitem{moore} G. D. Moore, {\tt hep-ph}/9810313.

\bibitem{blaizot} J. P. Blaizot and E. Iancu, Phys. Rev. Lett. 
{\bf 70}, 3376 (1993); Nucl. Phys. {\bf B417}, 608 (1994).

\bibitem{kelly} P. F. Kelly, Q. Liu, C. Lucchesi and C. Manuel, 
 Phys. Rev. Lett. {\bf 72}, 3461 (1994); Phys. Rev. D {\bf 50}, 
 4209 (1994). 
 
\bibitem{arnold} P. Arnold, D. T. Son and L. G. Yaffe, {\tt hep-ph}/9810216;
 {\tt hep-ph}/9901304.
 
\bibitem{blaizot2} J. P. Blaizot and E. Iancu, {\tt hep-ph}/9903389. 

\bibitem{selikhov} A. V. Selikhov, Phys. Lett. {\bf B268}, 263 (1991); 
 Phys. Lett. {\bf B285}, 398(E) (1992).
 
\bibitem{klim} Y. L. Klimontovich, {\sl Statistical Physics} (Harwood 
Academic Publishers, 1986).

\bibitem{lif} E. M. Lifshitz and L. P. Pitaevskii, {\sl Physical Kinetics} 
(Pergamon Press, Oxford, 1981). 

\bibitem{litim} D. F. Litim and C. Manuel, {\tt hep-ph}/9902430. 

\bibitem{sta} L. D. Landau and E. M. Lifshitz, {\sl Statistical Physics}, 
Part 1 (3rd edition),  (Pergamon Press, Oxford, 1980). 

\bibitem{sel2} A. V. Selikhov and M. Gyulassy, Phys. Lett. {\bf B316}, 
373 (1993).




\end{references}
\end{document}